\documentclass[pra,twocolumn,twoside,showpacs,footinbib,bibnotes]{revtex4}

\usepackage{graphicx}
\usepackage{amsmath,bm}

\newcommand{\eq}[1]{\begin{equation}#1\end{equation}}
\newcommand{\eqmulti}[1]{\begin{equation}\begin{split}#1\end{split}\end{equation}}
\newcommand{\eqgather}[1]{\begin{gather}#1\end{gather}}

\newcommand{\ket}[1]{\ensuremath{\,|{#1}\rangle}}
\newcommand{\braket}[2]{\ensuremath{\langle{#1}|{#2}\rangle}}
\newcommand{\matrixe}[3]{\ensuremath{\langle{#1}|\,{#2}\,|{#3}\rangle}}

\newcommand{\op}[1]{\ensuremath{\bm{#1}}}
\newcommand{\dd}{\ensuremath{\mathrm{d}}}
\newcommand{\ddd}[2][]{\ensuremath{\dd^{#1}#2\;}}

\newcommand{\rO}{\ensuremath{\op{r}}}

\newcommand{\HO}{\ensuremath{\op{H}}}

\newcommand{\xV}{\ensuremath{\vec{x}}}

\newcommand{\pOV}{\ensuremath{\vec{\op{p}}}}

\newcommand{\xOV}{\ensuremath{\vec{\op{x}}}}

\newcommand{\EC}{\ensuremath{\mathcal{E}}}

\newcommand{\nablaV}{\ensuremath{\vec{\nabla}}}

\newcommand{\nB}{\ensuremath{n_{\text{B}}}}
\newcommand{\nF}{\ensuremath{n_{\text{F}}}}

\newcommand{\PhiB}{\ensuremath{\Phi_{\text{B}}}}

\newcommand{\aB}{\ensuremath{a_{\text{B}}}}
\newcommand{\aBF}{\ensuremath{a_{\text{BF}}}}

\newcommand{\B}{\ensuremath{\text{B}}}
\newcommand{\F}{\ensuremath{\text{F}}}

\newcommand{\BF}{\ensuremath{\text{BF}}}

\begin{document}

\title{Mean-field instability of trapped dilute boson-fermion mixtures}

\author{R. Roth}
\email{r.roth@gsi.de}
%\homepage[\\ URL: ]{http://theory.gsi.de/~trap/}

\author{H. Feldmeier}
\email{h.feldmeier@gsi.de}

\affiliation{Gesellschaft f\"ur Schwerionenforschung (GSI), 
  Planckstr. 1, 64291 Darmstadt, Germany}

\date{\today}

\begin{abstract}
The influence of boson-boson and boson-fermion interactions on the
stability of a binary mixture of bosonic and fermionic atoms is
investigated. The density profiles of the trapped mixture are obtained
from direct numerical solution of a modified Gross-Pitaevskii equation
that is self-consistently coupled to the mean-field generated by the
interaction with the fermionic species. The fermions which in turn
feel the mean-field created by the bosons are treated in Thomas-Fermi
approximation. We study the effects of different combinations of signs
of the boson-boson and the boson-fermion scattering lengths and
determine explicit expressions for critical particle numbers as
function of these scattering lengths.
\end{abstract}

% 03.75.-b Matter waves
% 03.75.Fi Phase coherent atomic ensembles; quantum condensation
%          phenomena
% 32.80.Pj Optical cooling of atoms; trapping

\maketitle

%%%%%%%%%%%%%%%%%%%%%%%%%%%%%%%%%%%%%%%%%%%%%%%%%%%%%%%%%%%%%%%%%%
%%%%%% introduction %%%%%%%%%%%%%%%%%%%%%%%%%%%%%%%%%%%%%%%%%%%%%%
%%%%%%%%%%%%%%%%%%%%%%%%%%%%%%%%%%%%%%%%%%%%%%%%%%%%%%%%%%%%%%%%%%

Recent experimental successes in the trapping and cooling of mixtures
of bosonic and fermionic atoms \cite{TrSt01,ScKh01,GoPa01} constitute
a new branch in the field of trapped ultracold gases.  Similar to the
purely bosonic gases boson-fermion mixtures offer unique possibilities
to study fundamental quantum phenomena. Moreover they appear as a
promising candidate to realize a BCS transition to a superfluid phase
of the fermionic component \cite{HoFe97}. One of the most appealing
features of these systems is that the strength of the interaction
between the atoms can be tuned in a wide range by utilizing a Feshbach
resonance \cite{HoSt98}.

For the sympathetic cooling of a Fermi gas in binary boson-fermion
mixtures the collapse caused by attractive interactions is responsible
for a severe limitation of the lowest achievable temperature
\cite{TrSt01,ScKh01}.  The occurrence of a mean-field instability when
the density or particle number exceeds a critical value was already
studied experimentally for purely bosonic systems \cite{DoCl01}. In
this communication we investigate the interplay between boson-boson
and boson-fermion interactions and the implications for the stability
of the mixture.

%%%%%%%%%%%%%%%%%%%%%%%%%%%%%%%%%%%%%%%%%%%%%%%%%%%%%%%%%%%%%%%%%%
%%%%%% energy functional %%%%%%%%%%%%%%%%%%%%%%%%%%%%%%%%%%%%%%%%%
%%%%%%%%%%%%%%%%%%%%%%%%%%%%%%%%%%%%%%%%%%%%%%%%%%%%%%%%%%%%%%%%%%

In order to describe the properties of the binary boson-fermion
mixture at zero temperature we first construct the energy functional
in mean-field approximation. The many-body state $\ket{\Psi}$ of the
mixed boson-fermion system is a direct product of a symmetric
$N_{\B}$-body state $\ket{\Psi_{\B}}$ for the bosonic species and an
antisymmetric $N_{\F}$-body state $\ket{\Psi_{\F}}$ for the
fermions. The Hamiltonian of the interacting mixture reads
\eqmulti{ \label{eq:hamiltonian}
  \HO 
  &= \sum_{i=1}^{N_{\B}}\! 
    \bigg[\frac{\pOV_i^2}{2 m_{\B}} + U_{\B}(\xOV_i)\bigg]  
  + \sum_{i=N_{\B}+1}^{N_{\B}+N_{\F}}\! 
    \bigg[\frac{\pOV_i^2}{2 m_{\F}} + U_{\F}(\xOV_i)\bigg] \\
  &+\! \sum_{i<j=1}^{N_{\B}}\!\! 
    \frac{4\pi\aB}{m_{\B}}\, \delta^{(3)}(\rO_{ij})
  + \sum_{i=1}^{N_{\B}} \sum_{j=N_{\B}+1}^{N_{\B}+N_{\F}}\!
    \frac{4\pi\aBF}{m_{\BF}}\, \delta^{(3)}(\rO_{ij}) .  
}
The first line contains the kinetic energy operators and the external
trapping potentials $U_{\B}(\xV)$ and $U_{\F}(\xV)$ for the bosonic
and the fermionic species with masses $m_{\B}$ and $m_{\F}$,
respectively.

Due to the large average distance the atom-atom interaction can in
general be described by an effective contact interaction for all
partial waves \cite{RoFe01a,RoFe00}. In a binary boson-fermion mixture
one has to distinguish three interaction types: boson-boson
interactions, boson-fermion interactions, and fermion-fermion
interactions. The s-wave interaction between two bosons is described
by the third term of Eq. \eqref{eq:hamiltonian}, where $a_{\B}$ is the
s-wave scattering length. Since we consider a pure Bose-Einstein
condensate at zero temperature only the s-wave term is needed. Higher
even partial waves are negligible. For the interaction between a
bosonic and a fermionic atom s- and p-wave terms contribute. The
operator of the s-wave boson-fermion contact interaction forms the
last term of the Hamiltonian \eqref{eq:hamiltonian}, where $\aBF$ is
the corresponding s-wave scattering length and $m_{\BF} = 2 m_{\B}
m_{\F}/(m_{\B}+m_{\F})$ is twice the reduced mass of the pair.  Since
the s-wave interaction dominates in many cases of interest, e.g. the
${}^6$Li/${}^7$Li mixture, we will neglect the p-wave interaction for
this discussion. For identical fermions s-wave contact interactions
are prohibited by the Pauli principle. The first nonvanishing
contribution is the nonlocal p-wave interaction, which will also be
neglected in the following. However, p-wave interactions can have
significant influence on the structure and stability of the fermionic
component as we discussed earlier \cite{RoFe01a, RoFe01c}.

%%%%%%%%%%%%%%%%%%%%%%%%%%%%%%%%%%%%%%%%%%%%%%%%%%%%%%%%%%%%%%%%%%
%%%%% Energy density %%%%%%%%%%%%%%%%%%%%%%%%%%%%%%%%%%%%%%%%%%%%%
%%%%%%%%%%%%%%%%%%%%%%%%%%%%%%%%%%%%%%%%%%%%%%%%%%%%%%%%%%%%%%%%%%
%%
%%%%% figure %%%%%%%%%%%%%%%%%%%%%%%%%%%%%%%%%%%%%%%%%%%%%%%%%%%%%
\begin{figure*}
\begin{minipage}{0.72\textwidth}
\includegraphics[width=1\textwidth]{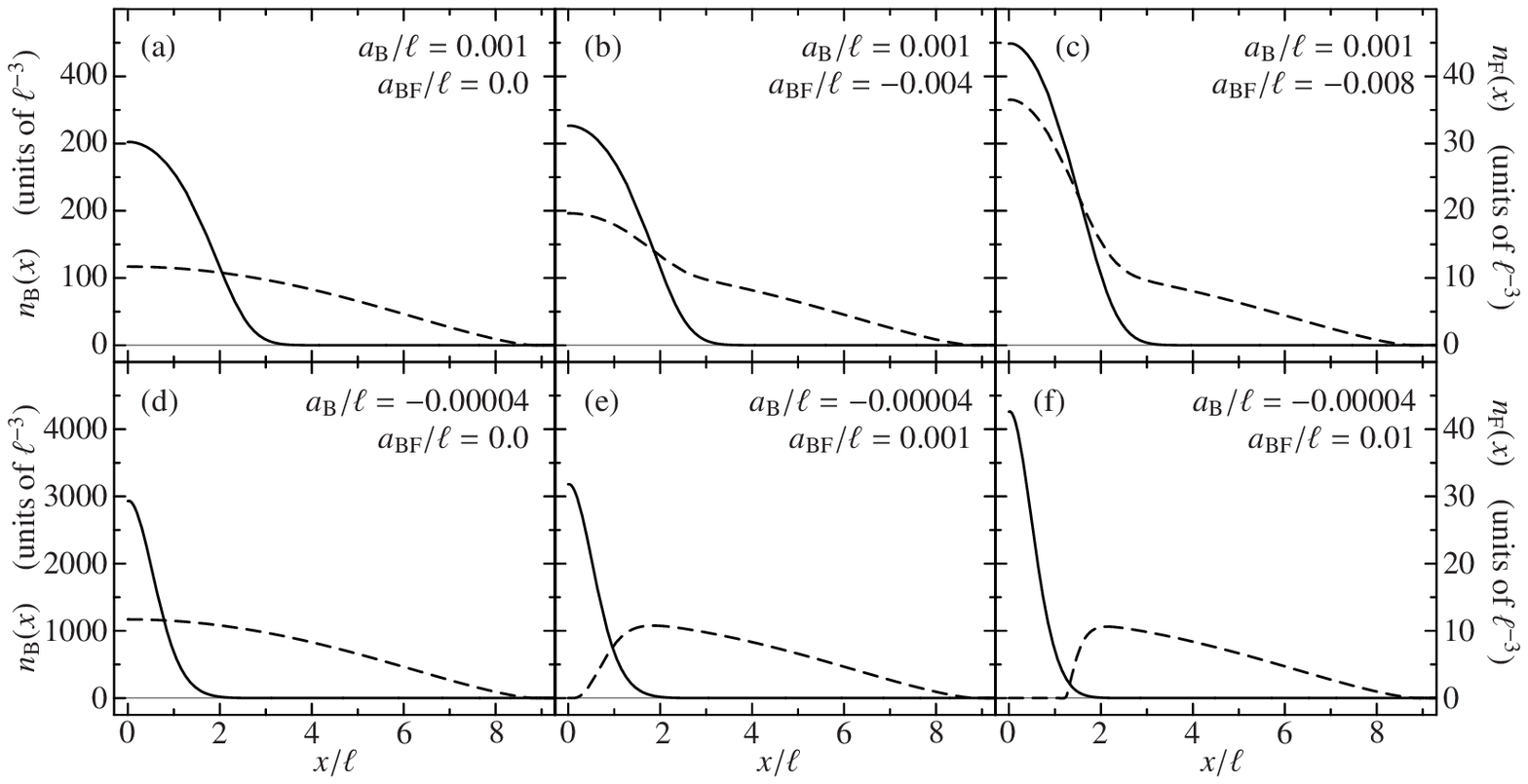}
\end{minipage}\hfill
\begin{minipage}{0.25\textwidth}
\caption{Radial density profiles of a boson-fermion mixture with
  $N_{\F}=N_{\B}=10\,000$ for different interaction strengths. The
  boson density $\nB(\xV)$ is given by the solid line (left scale) and
  the fermion density $\nF(\xV)$ by the dashed line (right scale). The
  upper row shows examples with increasing boson-fermion attraction
  and fixed $\aB/\ell=0.001$. The lower row depicts examples with
  increasing boson-fermion repulsion and $\aB/\ell=-0.00004$.}
\label{fig:density_profiles}
\end{minipage}
\end{figure*}
%%%%%%%%%%%%%%%%%%%%%%%%%%%%%%%%%%%%%%%%%%%%%%%%%%%%%%%%%%%%%%%%%%

The expectation value of the Hamiltonian \eqref{eq:hamiltonian}
calculated with the many-body state
$\ket{\Psi}=\ket{\Psi_{\B}}\otimes\ket{\Psi_{\F}}$, defines the energy
density of the mixture
\eq{ \label{eq:energy}
  E = \matrixe{\Psi}{\HO}{\Psi}
  = \int\ddd[3]{x} \EC[\nB,\nF](\xV) .
}
The energy density $\EC[\nB,\nF]$ is decomposed into a purely bosonic
part ($\B$), a fermionic part ($\F$) and the interaction part between
the two species ($\BF$)
\eq{
  \EC[\nB,\nF]
  = \EC_{\B}[\nB] + \EC_{\F}[\nF] +  \EC_{\BF}[\nB,\nF] .
}

For the calculation of $\EC_{\B}[\nB]$ we assume that the bosons are
in a pure Bose-Einstein condensate, i.e. the $N_{\B}$-boson state
$\ket{\Psi_{\B}}$ is given by a direct product of identical single
particle states $\ket{\phi_{\B}}$. This immediately leads to the
standard Gross-Pitaevskii energy density \cite{DaGi99}
\eqmulti{
  \EC_{\B}[\nB](\xV) 
  &=  U_{\B}(\xV)\, \nB(\xV) + \frac{1}{2m_{\B}} |\nablaV \nB^{1/2}(\xV)|^2 \\
  &+ \frac{2\pi\,\aB}{m_{\B}}\, \nB^2(\xV) ,
}
where $\nB(\xV) = \Phi_{\B}^2(\xV) = N_{\B} \braket{\xV}{\phi_{\B}}^2$
is the ground-state boson density.

For the evaluation of the parts of the energy density that involve the
fermionic species we employ the Thomas-Fermi approximation. It was
shown that this is an excellent approximation for the particle numbers
considered here \cite{NyMo99}. The fermionic part of the energy
density in Thomas-Fermi approximation reads \cite{RoFe01a}
\eq{
  \EC_{\F}[\nF](\xV) 
  = U_{\F}(\xV)\, \nF(\xV) 
  + \frac{3^{5/3}\, \pi^{4/3}}{5\,2^{1/3}\,m_{\F}}\; \nF^{5/3}(\xV).
}
The interaction between the bosonic and the fermionic species
yields the contribution
\eq{
  \EC_{\BF}[\nB,\nF](\xV)
  = \frac{4\pi\,\aBF}{m_{\BF}}\; \nB(\xV)\, \nF(\xV) . 
}

The functional minimization of the total energy \eqref{eq:energy}
under the constraint of given numbers of bosons and fermions leads to
the density profiles of the trapped gas. The constraints are
implemented by introducing the chemical potentials $\mu_{\B}$ and
$\mu_{\F}$ and minimizing the transformed energy
\eq{ \label{eq:transf_energy}
  F 
  = \int\ddd[3]{x} \big[ \EC[\nB,\nF](\xV) - \mu_{\B} \nB(\xV) 
    - \mu_{\F} \nF(\xV) \big] .
}
Variation with respect to the fermion density $\nF(\xV)$ immediately
leads to an equation for the density distribution
\eq{ \label{eq:fermiondensity}
  \nF(\xV) 
  = \frac{\sqrt{2 m_{\F}^3}}{3 \pi^2} \bigg[\mu_{\F} - U_{\F}(\xV) 
    - \frac{4\pi \aBF}{m_{\BF}} \nB(\xV) \bigg]^{3/2} .
}
From the functional variation of the energy \eqref{eq:transf_energy}
with respect to the bosonic density we obtain an Euler-Lagrange
differential equation for $\PhiB(\xV) = \sqrt{\nB(\xV)}$
\eqmulti{ \label{eq:grosspitaevskii}
  \bigg[-\frac{1}{2m_{\B}}\nablaV^2 &+ U_{\B}(\xV) 
    + \frac{4\pi \aBF}{m_{\BF}} \nF(\xV) \\ 
  &+ \frac{4\pi \aB}{m_{\B}}
    \PhiB^2(\xV) \bigg] \PhiB(\xV) = \mu_{\B} \PhiB(\xV) .
}
This is a modified Gross-Pitaevskii equation which includes the
mean-field contribution generated by the interaction with the
fermionic species.

The simultaneous solution of the coupled
Eqs. \eqref{eq:fermiondensity} and \eqref{eq:grosspitaevskii} gives
the density profiles of the two species. We solve the nonlinear
differential equation \eqref{eq:grosspitaevskii} with an efficient
quantum diffusion algorithm using a fast Fourier transformation
\cite{Feag94}. The ground state solution is extracted by successive
application of the time evolution operator for a small imaginary time
step. We accomplish the solution of the coupled problem by a simple
iterative procedure: \textcircled{\small0} the boson density is
initialized with the Gaussian profile of the noninteracting Bose gas
with $N_{\B}$ particles. \textcircled{\small1} The fermion density is
calculated using \eqref{eq:fermiondensity} with $\mu_{\F}$ adjusted
such that the integral over $\nF(\xV)$ gives the desired particle
number $N_{\F}$. \textcircled{\small2} A single imaginary time-step is
performed using the mean-field according to the boson and fermion
densities obtained in the previous two steps.  The resulting
$\nB(\xV)$ is normalized to $N_{\B}$ and used as initialization for
the next iteration cycle.
 
With these tools we investigate the instability of the boson-fermion
mixture against collapse induced by attractive boson-boson or
fermion-boson interactions. In order to keep the discussion simple we
restrict ourselves to spherical symmetric systems with equal
numbers of bosons and fermions $N_{\B} = N_{\F}$. We assume
equal masses for the two species $m=m_{\B} = m_{\F} = m_{\BF}$ and
identical parabolic trapping potentials $U_{\B}(x) = U_{\F}(x) =
x^2/(2m\ell^4)$. The oscillator length $\ell = (m \omega)^{-1/2}$
serves as fundamental length unit for the numerical treatment. A more
general treatment including asymmetric boson-fermion mixtures in
deformed traps will be presented in a subsequent paper.

%%%%%%%%%%%%%%%%%%%%%%%%%%%%%%%%%%%%%%%%%%%%%%%%%%%%%%%%%%%%%%%%%%
%%%%% B+ BF- %%%%%%%%%%%%%%%%%%%%%%%%%%%%%%%%%%%%%%%%%%%%%%%%%%%%%
%%%%%%%%%%%%%%%%%%%%%%%%%%%%%%%%%%%%%%%%%%%%%%%%%%%%%%%%%%%%%%%%%%

First we consider the case of repulsive boson-boson and attractive
boson-fermion interactions ($\aB\ge0, \aBF<0$). Here the attractive
interaction between the species induces a mean-field collapse if the
densities or particle numbers exceed a critical value. The upper row
of Fig. \ref{fig:density_profiles} shows the density profiles of
(meta)stable configurations with $N_{\B} = N_{\F} = 10^4$ particles
for three different values of the boson-fermion scattering length
$\aBF$ and fixed $\aB/\ell=0.001$ (corresponds to
$\aB\approx20\,a_{\text{Bohr}}$ for a typical trap with
$\ell=1\,\mu$m). Due to the Pauli principle the fermionic density
distribution is much more spread out and has a significantly lower
central density than the bosonic distribution with the same particle
number (notice the different scales for $\nB(\xV)$ and $\nF(\xV)$ in
Fig. \ref{fig:density_profiles}). Attractive boson-fermion
interactions generate an attractive mean-field for bosons proportional
to the density of the fermions and vice versa. This causes an increase
of both densities in the overlap region as can be seen in panel (b) of
Fig. \ref{fig:density_profiles}. With increasing strength of the
boson-fermion attraction the fermion density grows substantially. As
shown in Fig. \ref{fig:density_profiles}(c) the fermion density can
easily be increased by a factor $3$ compared to the noninteracting
case.

If the strength of the boson-fermion interaction exceeds a critical
value then the mixture collapses towards high densities. In this case
the attractive mean-field is not stabilized by the positive kinetic
energy contribution or the repulsive boson-boson interaction any more,
i.e. the gas can lower its energy by contracting and increasing the
density in the central region. This phenomenon was investigated in
detail for purely bosonic \cite{ElHu00} and purely fermionic
\cite{RoFe01a,RoFe01c} systems. To our knowledge the collapse of
boson-fermion mixtures was studied only inchoately using parameterized
density profiles \cite{MiSu01}.

We can determine the critical boson-fermion scattering length $\aBF$
for which the collapse occurs with a rather simple numerical
procedure. The quantum diffusion algorithm used to obtain the solution
of the Gross-Pitaevskii equation \eqref{eq:grosspitaevskii} diverges
if the mean-field instability occurs, i.e., the change of the boson
density in the trap center increases for successive imaginary time
steps. Thus by observing the convergence behavior of the central
density during the imaginary time evolution we can decide whether the
mixture is stable or collapses. 

To obtain a simple measure for the stability we proceed in two steps:
First we determine numerically the critical boson-fermion scattering
length for a set of particle numbers $N_{\F}=N_{\B} = 500\dots10^6$
and boson-boson scattering lengths $\aB/\ell=0\dots0.003$. Then we fit
a parametrization which connects the particle number with the two
scattering lengths to this data set. This leads to an expression for
the critical particle number $N_{\text{cr}}$ as function of the
scattering lengths $\aB/\ell\ge0$ and $\aBF/\ell<0$:
\eqgather{ \label{eq:Ncrit_b+bf-} 
  N_{\text{cr}}(\aB,\aBF,\ell)
  = \frac{0.283}{|\aBF/\ell|^{1.78}} 
    + \frac{0.374\,(\aB/\ell)^{2.36}}{|\aBF/\ell|^{5.69}}. 
}
Any mixture with $N_{\B}=N_{\F}>N_{\text{cr}}$ is unstable against
mean-field induced collapse.

\begin{figure}[t]
\includegraphics[width=0.95\columnwidth]{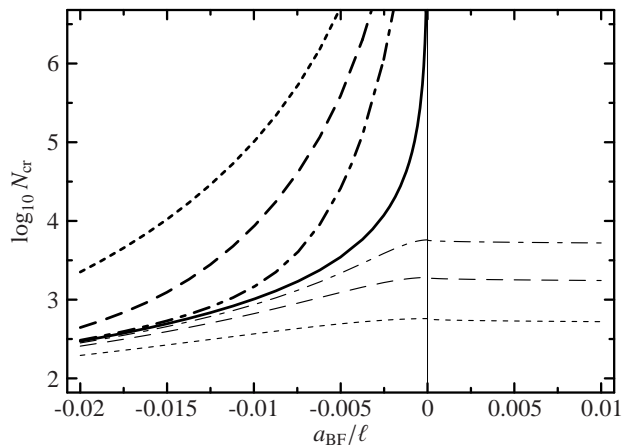}
\vskip-2ex
\caption{Logarithm of the critical particle number as function of 
  the boson-fermion scattering length $\aBF/\ell$ for different values
  of the boson-boson scattering length. The solid line corresponds to
  $\aB/\ell=0$. Thick lines show the behavior for repulsive
  boson-boson interactions: $\aB/\ell=0.0003$ (thick dash-dotted),
  $0.001$ (thick dashed), $0.003$ (thick dotted). Thin lines
  correspond to boson-boson attraction: $\aB/\ell=-0.0001$ (thin
  dash-dotted), $-0.0003$ (thin dashed), $-0.001$ (thin dotted).}
\label{fig:Ncrit}
\end{figure}

Figure \ref{fig:Ncrit} depicts the critical particle number as
function of the boson-fermion scattering length $\aBF$ for different
values of the boson-boson scattering length $\aB$. The thick lines
show the critical particle number obtained from \eqref{eq:Ncrit_b+bf-}
for repulsive boson-boson interactions. Notice that the values of
$\aB/\ell$ used in the plot are typically one order of magnitude
smaller than the range shown for $\aBF/\ell$; this emphasizes the
strong influence of the boson-boson interaction.
%%
%%%%% figure %%%%%%%%%%%%%%%%%%%%%%%%%%%%%%%%%%%%%%%%%%%%%%%%%%%%%
%%%%%%%%%%%%%%%%%%%%%%%%%%%%%%%%%%%%%%%%%%%%%%%%%%%%%%%%%%%%%%%%%%

We observe that a moderate boson-fermion attraction causes a severe
limitation of the particle number of the (meta)stable mixture. For
$\aB/\ell=0$ and $\aBF/\ell=-0.01$, which corresponds to
$\aBF\approx-200\,a_{\text{Bohr}}$ for $\ell=1\,\mu$m, the mixture is
stable only up to $N_{\B}=N_{\F}\approx1000$. The inclusion of a
repulsive boson-boson interaction leads to a significant
stabilization, i.e. an increase of the critical particle number. For
the above example, the critical particle number grows from
$N_{\text{cr}}\approx1000$ to $8500$ if a weak boson-boson repulsion
with $\aB/\ell=0.001$ is included.

%%%%%%%%%%%%%%%%%%%%%%%%%%%%%%%%%%%%%%%%%%%%%%%%%%%%%%%%%%%%%%%%%%
%%%%% B- BF- %%%%%%%%%%%%%%%%%%%%%%%%%%%%%%%%%%%%%%%%%%%%%%%%%%%%%
%%%%%%%%%%%%%%%%%%%%%%%%%%%%%%%%%%%%%%%%%%%%%%%%%%%%%%%%%%%%%%%%%%

As a second class of systems we consider mixtures where both, the
boson-fermion and the boson-boson interaction are attractive
($\aB<0,\,\aBF<0$). Compared to the previous class the attractive
boson-boson interaction enhances the attractive mean-field generated
by the boson-fermion interaction and abets the instability. The critical
particle numbers are significantly reduced.

We obtain a relation between the scattering lengths $\aB$ and $\aBF$
and the critical particle number in the same way as before. The
resulting parametrization of the critical particle number for $\aB<0$
and $\aBF<0$ reads
\eq{ \label{eq:Ncrit_b-bf-} 
  N_{\text{cr}}(\aB,\aBF,\ell) 
  = \frac{0.575}{|\aB/\ell| + 2.03 |\aBF/\ell|^{1.78}} .
}
For $\aBF=0$ we recover the well known relation between the critical
particle number and $|\aB|$ for a pure Bose gas \cite{DaGi99,ElHu00}.
The additional term in the denominator describes the modification of
$N_{\text{cr}}$ in the presence of an attractive boson-fermion
interaction. In the limiting case $\aB=0$ this relation coincides with
the corresponding limit of \eqref{eq:Ncrit_b+bf-}.

The thin curves in Fig. \ref{fig:Ncrit} show the behavior of the
critical particle number for $\aB/\ell<0$. The presence of the
attractive boson-boson interaction reduces the critical particle
number for $\aBF<0$ significantly. With increasing strength of the
boson-boson attraction the influence of the boson-fermion attraction
is largely reduced.

%%%%%%%%%%%%%%%%%%%%%%%%%%%%%%%%%%%%%%%%%%%%%%%%%%%%%%%%%%%%%%%%%%
%%%%% B- BF+ %%%%%%%%%%%%%%%%%%%%%%%%%%%%%%%%%%%%%%%%%%%%%%%%%%%%%
%%%%%%%%%%%%%%%%%%%%%%%%%%%%%%%%%%%%%%%%%%%%%%%%%%%%%%%%%%%%%%%%%%

Finally we consider mixtures with attractive boson-boson and repulsive
boson-fermion interactions ($\aB<0,\, \aBF\ge0$). The
${}^6$Li/${}^7$Li mixture used in the experiment of Truscott \emph{et
al.} \cite{TrSt01} belongs to this class of interactions.  The lower
row of Fig. \ref{fig:density_profiles} shows the density profiles for
three different values of $\aBF\ge0$. Already for a very weak
boson-fermion repulsion the two species separate spatially [see
Fig. \ref{fig:density_profiles}(e)], the bosons occupy the central
region of the trap (boson core) and the fermions constitute a shell
around it \cite{Molm98,NyMo99}. This structure may have interesting
implications for the mean-field instability of the bosons: The
fermionic shell compresses the boson core, i.e. increases the maximum
boson density as it is clearly seen in
Fig. \ref{fig:density_profiles}(f).  This could promote the mean-field
collapse in the presence of attractive boson-boson interactions and
lower the critical particle number.

The dependence of $N_{\text{cr}}$ on the scattering lengths, which is
obtained from the direct numerical solution of the coupled problem, is
for $\aB<0$ and $\aBF\ge0$ very well described by the parametrization
\eq{ \label{eq:Ncrit_b-bf+}
  N_{\text{cr}}(\aB,\aBF,\ell) 
  = \frac{0.575 - 0.230\, (\aBF/\ell)^{0.333}}{|\aB/\ell|} . 
}
The thin curves for $\aBF/\ell>0$ in Fig. \ref{fig:Ncrit} show this
dependence. Obviously the influence of the repulsive
boson-fermion interaction on the critical particle number is marginal.
The critical particle number reduces slightly if  $\aBF/\ell$ is
increased. This can be attributed to the compression of the boson core
mentioned before. Although the boson-fermion interaction
has a strong influence on the density profiles, its influence on the
critical particle number is negligible.

%%%%%%%%%%%%%%%%%%%%%%%%%%%%%%%%%%%%%%%%%%%%%%%%%%%%%%%%%%%%%%%%%%
%%%%% Summary %%%%%%%%%%%%%%%%%%%%%%%%%%%%%%%%%%%%%%%%%%%%%%%%%%%%
%%%%%%%%%%%%%%%%%%%%%%%%%%%%%%%%%%%%%%%%%%%%%%%%%%%%%%%%%%%%%%%%%%

In summary we have investigated the mean-field instability of binary
boson-fermion mixtures with equal particle numbers. We solved the
coupled Gross-Pitaevskii equation numerically and obtained the
critical particle number as function of the boson-boson and the
boson-fermion scattering lengths. We have shown that the boson-boson
and the boson-fermion interaction have very different effects on the
collapse: In the presence of attractive boson-fermion interactions
($\aBF<0$) the system is stabilized by weak repulsive and destabilized
by weak attractive boson-boson interactions. In contrast repulsive
boson-fermion interactions ($\aBF>0$) cannot stabilize the mixture
against collapse due to attractive boson-boson interactions.  With
regard to the recent experiments using mixtures of ${}^6$Li
($F=3/2,m_F=3/2$) and ${}^7$Li ($F=2,\,m_F=2$) \cite{TrSt01,ScKh01}
this implies that the collapse due to the attractive boson-boson
interaction is influenced only marginally by the boson-fermion
repulsion.

%%%%%%%%%%%%%%%%%%%%%%%%%%%%%%%%%%%%%%%%%%%%%%%%%%%%%%%%%%%%%%%%%%
%%%%%%%%%%%%%%%%%%%%%%%%%%%%%%%%%%%%%%%%%%%%%%%%%%%%%%%%%%%%%%%%%%
%%%%%%%%%%%%%%%%%%%%%%%%%%%%%%%%%%%%%%%%%%%%%%%%%%%%%%%%%%%%%%%%%%
%\bibliography{/u/rroth/tex/bib/bib_trap.bib}

%%%%%%%%%%%%%%%%%%%%%%%%%%%%%%%%%%%%%%%%%%%%%%%%%%%%%%%%%%%%%%%%%%
%%%%%%%%%%%%%%%%%%%%%%%%%%%%%%%%%%%%%%%%%%%%%%%%%%%%%%%%%%%%%%%%%%
%%%%%%%%%%%%%%%%%%%%%%%%%%%%%%%%%%%%%%%%%%%%%%%%%%%%%%%%%%%%%%%%%%

\end{document}